\documentclass[journal=nalefd, manuscript=letter]{achemso}

\usepackage{chemformula} 
\usepackage[T1]{fontenc} 
\usepackage{amsmath}

\author{Marlon R\"uck}
\email{marlon.rueck@tum.de}
\affiliation[TUM]
{Department of Electrical and Computer Engineering, Technical University Of Munich, 80333 M\"unchen, Germany }
\author{Aliaksandr Bandarenka}
\affiliation[TUM]
{Physics Department, Technical University Of Munich, 85748 Garching, Germany }
\author{Federico Calle-Vallejo}
\affiliation[]
{Department of Materials Science and Physical Chemistry, Institute of Theoretical and Computational Chemistry (IQTC), University of Barcelona, 08028 Barcelona, Spain}
\author{Alessio Gagliardi}
\affiliation[TUM]
{Department of Electrical and Computer Engineering, Technical University Of Munich, 80333 M\"unchen, Germany }

\title[]
  {Oxygen Reduction Reaction: Rapid Prediction of Mass Activity of Unstrained Nanostructured Platinum Electrocatalysts}

\abbreviations{IR,NMR,UV}

\begin{document}



\newpage
\begin{abstract}
  Tailored Pt nanoparticle catalysts are promising candidates to accelerate the oxygen reduction reaction (ORR) in fuel cells. However, the search for active nanoparticle catalysts is hindered by laborious effort of experimental synthesis and measurements. On the other hand, DFT-based approaches are still time consuming and often not efficient. In this study, we introduce a computational model which enables rapid catalytic activity calculation of unstrained pure Pt nanoparticle electrocatalysts. The generic setup of the computational model is based on DFT results and experimental data obtained worldwide over the past ca 20 years; whereas, importantly, the computational model dispenses with DFT calculations during runtime. This realizes feasible and sharply reduced computation effort in comparison to theoretical approaches where DFT calculations must be performed for each nanoparticle individually. Regarding particle size effects on Pt nanoparticles, experimental catalytic mass activities from previous studies are accurately reproduced by our computational model. Shedding light on the parameter space of particle size effects, this study enables predictions beyond available experiments: Our computational model identifies potential enhancement in mass activity up to 190\% over the experimentally detected maximum. Importantly, the rapid activity calculation enabled by our computational model may pave the way for extensive nanoparticle screening to expedite the search for improved electrocatalysts. \\ 
\end{abstract}

\begin{figure}[H]
\centering
\includegraphics[width=0.94\textwidth]{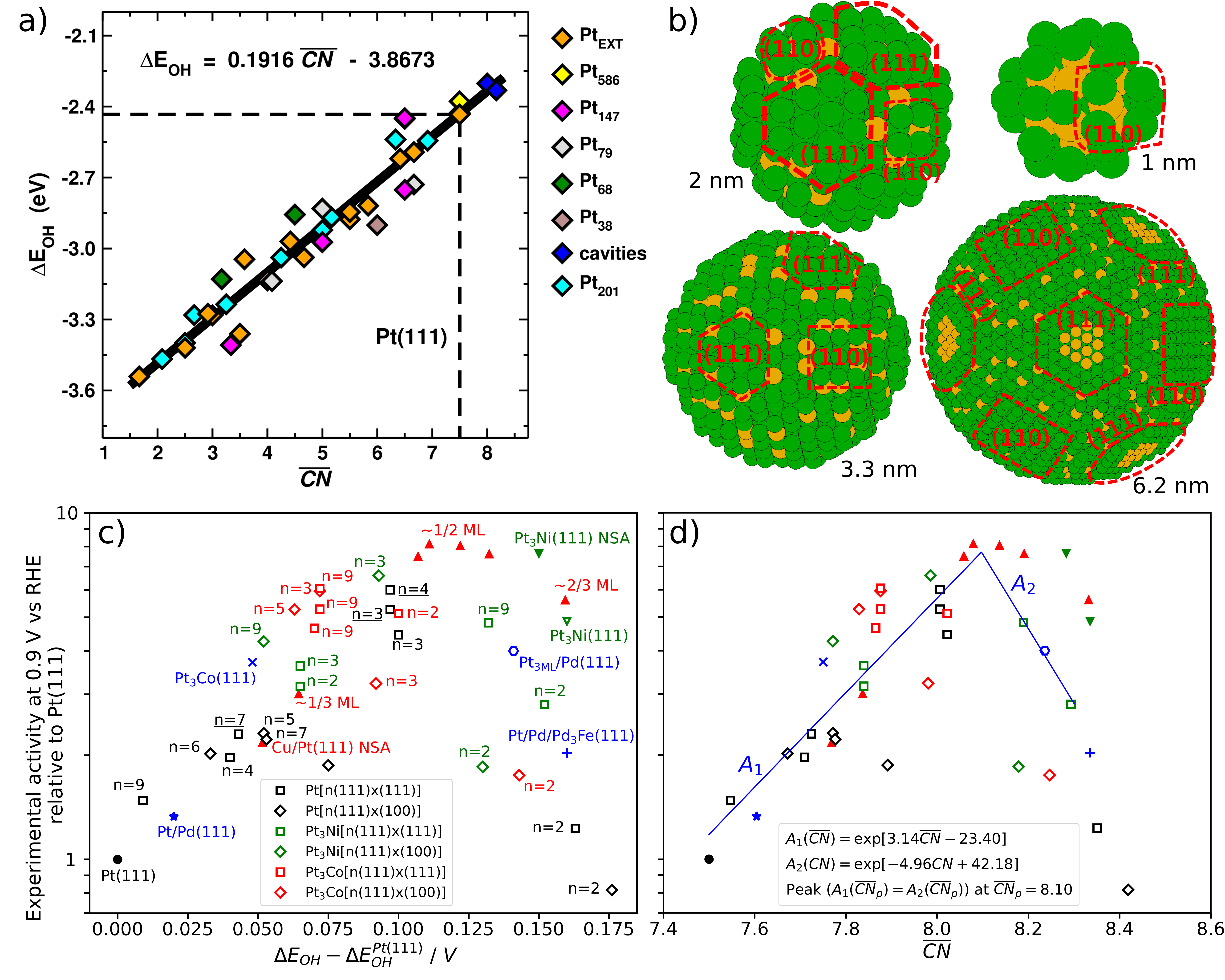}
\caption{(a) Adsorption energies $\Delta E_{OH}$ are calculated by DFT (see Supporting Information) on diversely coordinated sites in various catalyst shapes of different sizes; including tetrahedrons (green), cuboctahedrons (magenta), truncated octahedrons (yellow, cyan, grey, brown), extended surfaces (orange) and cavities (blue). The linear dependence on $\overline{CN}$ is described by the linear function provided in the inset. (b) Spherical nanoparticles (printed by ASE \cite{Larsen2017}) at diameters of 1 nm, 2 nm, 3.3 nm and 6.2 nm are exemplified as they are investigated in this study. Active sites with $7.5\leq\overline{CN}\leq8.3$ are highlighted in yellow. Prominent low index surfaces are enclosed by dashed lines. (c) Relative experimental activities of various Pt stepped surfaces (forming terrace widths of length $n$) and Pt alloy fcc(111) single-crystals are plotted vs experimental OH-binding energies: (black open squares) Pt stepped surfaces; (green open squares) Pt$_3$Ni stepped surfaces; (red open squares) Pt$_3$Co stepped surfaces; (red up-pointing traingle) Cu/Pt(111) NSAs with full and partial (1/3 ML, 1/2 ML, 2/3 ML) surface Cu content; (full green down-pointing triangle) Pt$_3$Ni(111) NSA; (open green down-pointing triangle) bulk Pt$_3$Ni(111); (blue star) one monolayer of Pt on Pd(111); (blue plus) monolayer of Pt on annealed Pd$_3$Fe(111) electrode with one segregated Pd layer; (open blue octahedron) three monolayers of Pt on Pd(111); (blue x) bulk Pt$_3$Co(111). Pt stepped surfaces, which are highlighted by underlined terrace widths of length $n=3$, $n=4$, $n=7$, are taken from Ref. \cite{Calle-Vallejo2017}. Remaining data is taken from Ref. \cite{Colic2016} and sources therein.  The catalytic activities are measured at 0.9 V vs RHE in 0.1 M HCLO$_4$. (d) The linear scaling relation in a) maps the experimental binding energy in c) onto $\overline{CN}$. Linear regression data of the increasing and decreasing activity functions $A_1$ and $A_2$, respectively, is provided in the inset.}
\label{fig:Figure1}
\end{figure}

\begin{figure}[H]
\centering
\includegraphics[width=0.5\textwidth]{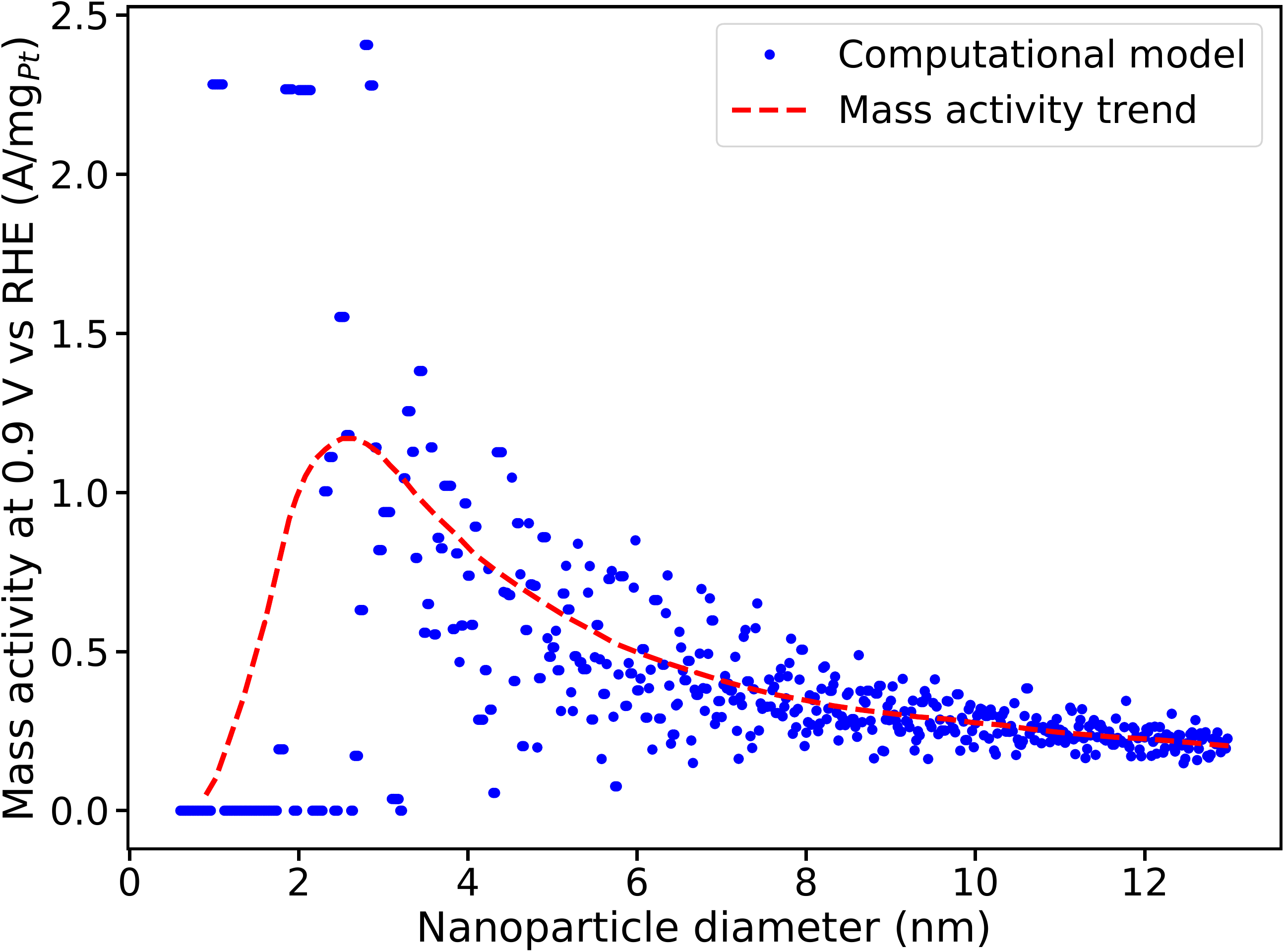}
\caption{ Computational results for the mass activity, which is obtained in absolut units of $A/mg_{Pt}$, versus nanoparticle sizes. 620 distinct nanoparticles, involving diameters between 0.6 nm and 13 nm in intervals of 0.02 nm, are taken into account. The dashed red curve depicts the overall mass activity trend which features a peak near 2.5 nm.}
\label{fig:Figure2}
\end{figure}

\begin{figure}[H]
\centering
\includegraphics[width=0.98\textwidth]{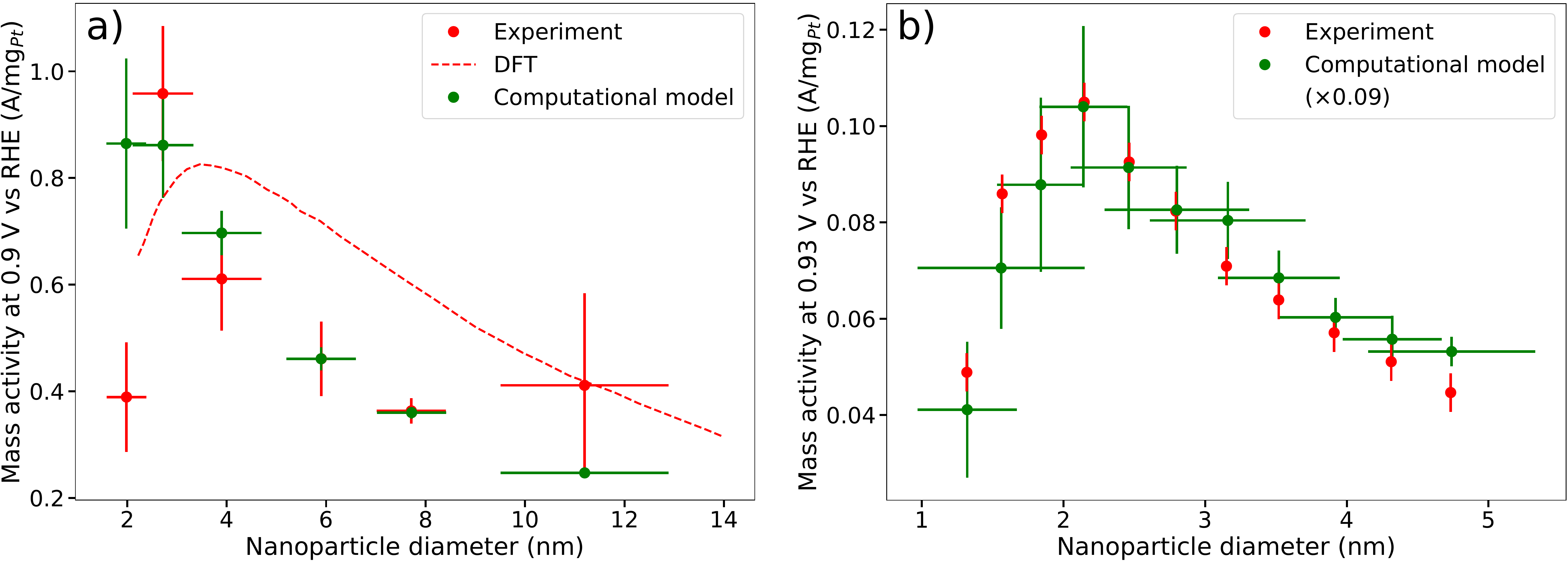}
\caption{ Experimental (red dots) and computational results (green dots) of the particle size effect. Vertical error bars for computational mass activities represent the standard error of the mean. (a) Experimental data is taken from {\it Perez-Alonso et al.} \cite{Perez-Alonso2012}. Computational mass activities are calculated and displayed in absolute units of $A/mg_{Pt}$. Diameter distributions employed in the computational model are adopted from the experimental study. (b) The experimental activities in this particular study from {\it Shao et al.} \cite{Shao2011g} differ from a) by one magnitude. Thus, unlike the absolute approach in a), all computational mass activities are multiplied by a factor of 0.09 as a fit to the experimental data. The standard deviation of the diameter is constrained between 0.18-0.35 nm similar to the experimental specification of 0.2-0.3 nm. 
}
\label{fig:Figure3}
\end{figure}

\begin{figure}[H]
\centering
\includegraphics[width=0.5\textwidth]{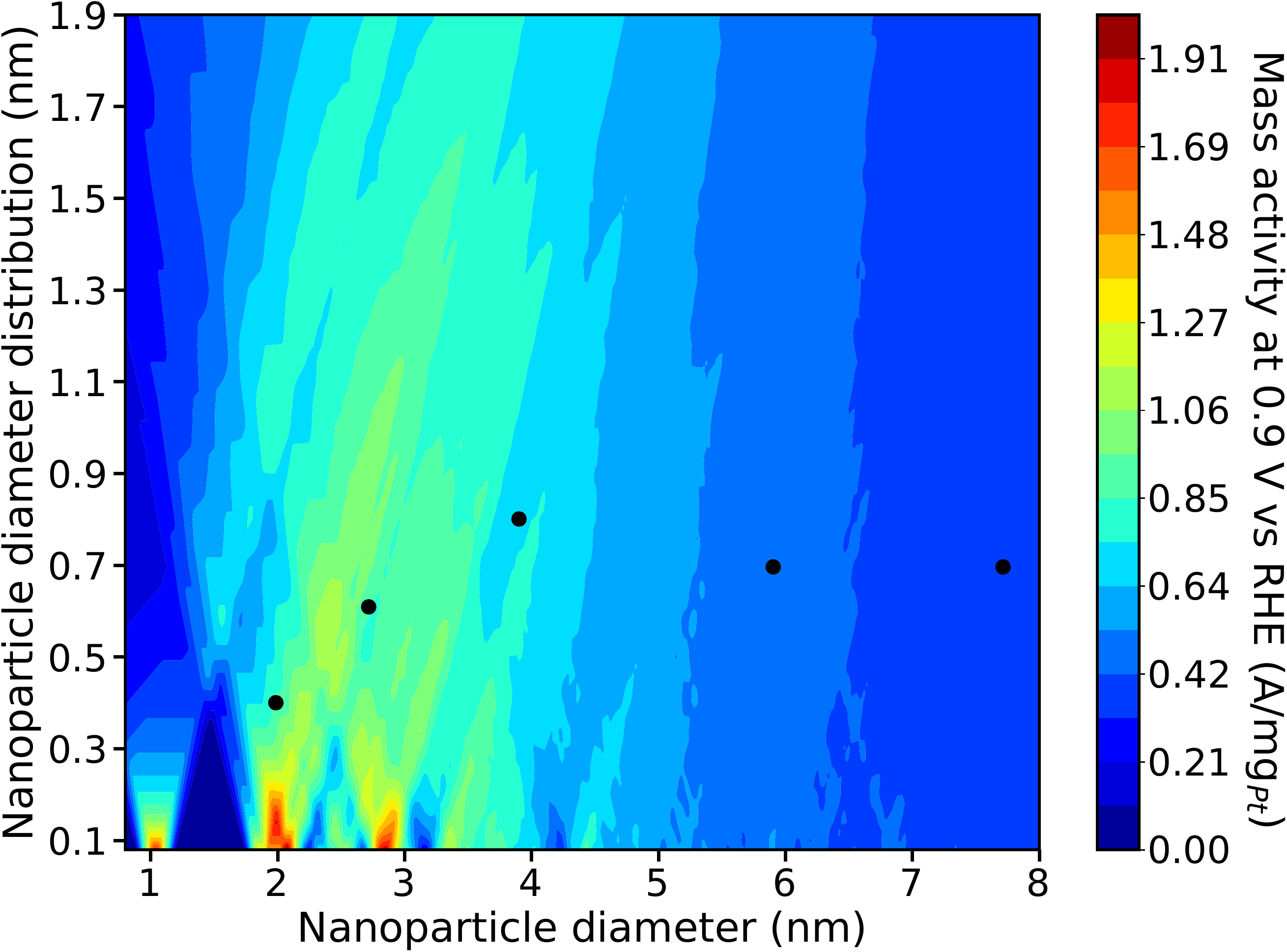}
\caption{ The contour plot elucidates the full parameter space of the particle size effect: Nanoparticle diameters (on the horizontal axis) and associated diameter distributions (on the vertical axis) are mapped onto the catalytic mass activity (presented by the color bar) in absolute units of $A/mg_{Pt}$. The experimental data from Figure \ref{fig:Figure3}a (labeled by black dots) is included. The contour plot reveals that highest mass activities (indicated by red colored areas) are harbored by nanoparticles at diameters of 1 nm, 2 nm and 2.9 nm with diameter distributions below 0.2 nm.}
\label{fig:Figure4}
\end{figure}

\newpage
Proton-exchange membrane fuel cells (PEMFCs) are suitable devices for versatile stationary and portable energy solutions \cite{Sharaf2014c}. Apart from industrial concerns on the durability of fuel cells \cite{Wu2008}, widespread commercialization of fuel cell technologies is impeded by the high costs of platinum which is required for the oxygen reduction reaction (ORR) at the cathode side of these devices \cite{Gasteiger2005c}. Due to different adsorption energies of reaction intermediates, the catalytic activity strongly depends on the catalyst surface structure \cite{C1CP21687B}. Structural sensitivity has been extensively studied on stepped surfaces which harbor considerably increased catalytic activities relative to Pt(111) \cite{Macia2004,Hitotsuyanagi2012,Colic2016}. Nanoparticle catalysts combine high surface to volume ratio with the capability to tailor active catalyst surface structures. Prominently, Pt nanowires have recently been fabricated at laboratory level which exceed the mass activity of current state-of-the-art commercial platinum on carbon supported (Pt/C) catalysts by a factor of 52 \cite{Li2016,Erfye2016}. Nevertheless, the progress in search for promising nanoparticle catalysts is restricted by complex synthesis on the experimental side and approaches solely based on expensive atomistic density functional theory (DFT) on the theory side \cite{Vines2014}. Suitable descriptors significantly promote the classification of catalyst structures into promising and inactive. From an early stage, d-band centers \cite{HAMMER200071} have been an important concept in this field. However, the necessity of DFT calculations for d-band studies further stimulated the identification of more affordable descriptors. To this end, conventional coordination numbers, counting the number of first nearest neighbors of the active sites, yield appropriate scaling relations for catalytic activities of extended surfaces \cite{Li2014,Calle-Vallejo2015a}. However, activity trends for nanoparticle catalysts remain out of scope due to finite size effects \cite{Kleis2011,Mpourmpakis2010}. \\
In recent studies, {\it Calle-Vallejo et al.} extended the concept of coordination number to the second nearest neighbors by means of generalized coordination numbers (gCN)
\begin{equation}
  \overline{CN}(i) = \sum_{j=1}^{n_i} \frac{cn(j)}{cn_{max}} ~.
  \label{eqn:gCN}
\end{equation}
Those have been proven simple descriptors for catalytic activities of various reactions \cite{Calle-Vallejo2014g,Calle-Vallejo2015,Calle-Vallejo2017,Strasser2018}. The conventional coordination numbers $cn(j)$ are summed up for all $n_i$ nearest neighbor sites $j$ of site $i$ such that finite size effects are considered explicitly. The maximal coordination number $cn_{max}$ (i.e. $cn_{max}=12$ for top sites in the fcc structure) yields a normalization to bulk atoms which are represented by $\overline{CN}=12$ equivalently to the conventional bulk coordination number. The Sabatier analysis for the oxygen reduction reaction, in which competing adsorption energies of the intermediates $^*$OH and $^*$OOH are evaluated similarly to an earlier study \cite{Norskov2004}, revealed an optimal adsorption energy tradeoff in range of $7.5 < \overline{CN}\leq 8.3$ where enhanced catalytic activity relative to Pt(111) is expected \cite{Calle-Vallejo2015,Calle-Vallejo2017}. Moreover, the DFT based adsorption energies of all crucial ORR intermediates, namely $^*$O, $^*$O$_2$, $^*$OH, and $^*$OOH, are linearly related with $\overline{CN}$ \cite{Calle-Vallejo2014g,Calle-Vallejo2015}. \\
On the experimental side, the adsorption potentials of $^*$OH can be obtained at model extended surfaces with respect to Pt(111) by the analysis of cyclic voltammograms \cite{Calle-Vallejo2017,Colic2016}. Furthermore, catalytic activities can be expressed by kinetic current densities at certain important electrode potentials \cite{Stamenkovic2007,Hitotsuyanagi2012,Shao2011g,Perez-Alonso2012} or using quasi exchange current densities within the Tafel approximation \cite{Macia2004}. It is observed that a weakening of OH-adsorption potentials with respect to Pt(111) up to $\sim0.1-0.15V$ results in larger current densities \cite{C4CP00260A,Calle-Vallejo2015,Calle-Vallejo2017,Stephens2012}. \\
Herein, we combine theoretical and experimental data to develop a computational model which calculates the catalytic activity of pure unstrained Pt nanoparticle electrocatalysts in a short computation time. We calculate OH-adsorption energies with reference to OH in the gas phase on diversely coordinated sites in multifaceted nanocatalyst shapes (as tetrahedrons, cuboctahedrons, truncated octahedrons and extended surfaces) by means of DFT. Note, however, that the difference between two adsorption energies does not depend on the gas-phase reference used, as long as the reference is identical. For details on the DFT calculations we refer the reader to our  Supporting Information. \\
The evaluation of coordination for all involved sites exposes fundamental linear dependence between OH-adsorption energies and $\overline{CN}$, which we present in Figure \ref{fig:Figure1}a. In addition, we use experimental data from literature \cite{Calle-Vallejo2017,Colic2016}, comprising catalytic activities versus experimentally observed OH-binding energies for extended Pt surfaces and Pt alloys, to draw the volcano plot in Figure \ref{fig:Figure1}c. Importantly, the above discussed linear relation from Figure \ref{fig:Figure1}a is then employed to map the experimental OH-binding energies from Figure \ref{fig:Figure1}c to an equivalent of $\overline{CN}$. The procedure is outlined in the Supporting Information. \\
As shown in Figure \ref{fig:Figure1}c, Pt alloys and pure Pt surfaces follow the same activity trends. Hence, one should notice that there is no discrepancy between the fact that Pt alloys and pure Pt surfaces are used to construct the resulting volcano plot in Figure \ref{fig:Figure1}d. The associated volcano-shaped catalytic activity trend agrees well with the aforementioned Sabatier analysis \cite{Calle-Vallejo2015} where enhanced catalytic activities relative to Pt(111) are expected for sites with generalized coordination $7.5 < \overline{CN}\leq 8.3$. The trend is captured by fit functions $A_1$ and $A_2$ (see Supporting Information) which form the peak of the volcano at $\overline{CN} = 8.1$. This corresponds to an OH-binding potential relative to Pt(111) of $\Delta E_{OH}-\Delta E^{Pt(111)}\approx 0.115~V$. Thus, unstrained Pt nanoparticle catalysts can be examined by evaluation of gCNs at all nanoparticle sites. To this end, the activity contributions of all sites are summed up according to the trend in Figure \ref{fig:Figure1}d. We discuss this essential step in more detail in the Supporting Information. Even more intriguingly, we exploit additional geometrical considerations and the total number of sites (which is pointed out in the Supporting Information) in order to yield mass activities not only relative to Pt(111), but rather in units of Amperes per milligram of Pt. Beyond the peak of the volcano at larger $\overline{CN}$ in Figure \ref{fig:Figure1}d, the activity trend is widely dispersed. At small $\overline{CN}$, undercoordinated sites may be affected by oxygenated species which leads to blocked catalytic processes at these centers \cite{Pohl2016}. Therefore, the activity contribution of sites with $\overline{CN} < 7.5$ or $\overline{CN} > 8.3$ is set to zero in our computational model, but all nanoparticle sites are taken into account for the mass activity prediction. It is noteworthy that our computational model does not employ any additional assumptions than those general considerations discussed above. \\
In this study, the nanoparticle catalysts are modeled by quasi-spherical shapes such as those exemplified in Figure \ref{fig:Figure1}b. Additional spherical nanoparticle catalysts comprising a broader range of diameters are presented in Figure \ref{fig:FigureS1} in the Supporting Information. Within our computational model, the mass activities for 620 distinct nanoparticles are evaluated for diameters ranging from 0.6 nm to 13 nm in small-scale 0.02 nm intervals. As it becomes apparent in Figure \ref{fig:Figure2}, the mass activity depends sensitively on the nanoparticle diameter and the overall mass activity trend features a peak near 2.5 nm. Thus, the analysis of Figure \ref{fig:Figure2} also shows that the size distribution of the nanoparticles turns out to be crucial for accurate catalytic activity prediction. Therefore, activities for distinct nanoparticle diameters are obtained by the mean activity within the diameter distribution. \\
The applicability of the computational model is further compared with experimental data. {\it Perez-Alonso et al.} and {\it Shao et al.} independently investigated nanoparticle size effects on the catalytic activity as shown in Figure \ref{fig:Figure3}a and Figure \ref{fig:Figure3}b, respectively \cite{Perez-Alonso2012,Shao2011g}. Note that the maximal mass activity has been detected at nanoparticle sizes between 2-3 nm which coincides with related experimental \cite{Leontyev2011,Gasteiger2005c} and theoretical \cite{Tritsaris2011,Tripkovic2014} studies. For the dataset in Figure \ref{fig:Figure3}a, the experimental diameter distribution is specified individually for each nanoparticle. We equally adapt the experimental diameter distributions in our computational model. Interestingly, the experimental mass activity trend is precisely reproduced by our computational approach. Furthermore, particularly regarding absolute units the computational and experimental mass activities coincide as the associated error intervals overlap; except for the smallest diameter near 2 nm where corrosion effects are believed to have degraded the experimental nanoparticle structure \cite{Perez-Alonso2012}. In this regard, it is important to emphasize that slight deviations in the size distribution may considerably affect the associated mass activity around diameters of 2 nm. By contrast, the second experimental dataset in Figure \ref{fig:Figure3}b comprises significantly lower mass activities at a level of $10\%$ compared to the mass activities in Figure \ref{fig:Figure3}a. Consequently, unlike the absolute approach in Figure \ref{fig:Figure3}a, the computational activity trend for the second experimental dataset in Figure \ref{fig:Figure3}b is scaled to fit the corresponding experimental trend. Multiplying all computational values by a scale factor of $0.09$ yields the best agreement with these particular experiments. Within the computational model, the standard deviation of the diameter is constrained between 0.18 nm and 0.35 nm. For the experimental measurements, the overall standard deviation is stated to be similarly between 0.2 nm and 0.3 nm \cite{Shao2011g}. As remarkable result, the computational trend is in good agreement with the experimental values. \\ 
Furthermore, it is important to note that the steep decrease in experimental activity around 2 nm in Figure \ref{fig:Figure3}a is considerably less pronounced in Figure \ref{fig:Figure3}b. This substantiates the assumption that corrosion has affected the surface structure of the 2 nm nanoparticles for the case shown in Figure \ref{fig:Figure3}a. \\
{\it Perez-Alonso et al.} compared the experimental results with an earlier theoretical study \cite{Tritsaris2011,Stephens2012} which is represented by the dashed curve in Figure \ref{fig:Figure3}a. Therein, nanoparticles are constructed by edged surface facets which differ from spherical shapes. Relative activities in arbitrary units are obtained via adsorption free energies from DFT calculations. The experimental trend is adequately captured in the sense that the mass activity peak at 2-4 nm is reproduced which is followed by a slightly flattened decrease in mass activity towards large diameters. However, the precision in nanoparticle size has not been taken into account in this DFT approach. Consequently, experimental and theoretical approaches still need to be brought in quantitative agreement. Remarkably, this has been achieved in the present computational model by explicit consideration of size distribution and absolute units, which constitutes a step forward compared to previous studies. \\
Furthermore, such quantitative agreement with experimental data ascertains that spherical nanoparticles serve as the appropriate model structures in order to simulate real nanoparticle catalysts. \\
The activity analysis of our computational model enables interesting nanoparticle size predictions with enhanced activity performance. Exploring the nanoparticle size effect at the maximum level of detail, we produced the contour plot in Figure \ref{fig:Figure4}a where the nanoparticle diameter range and associated diameter distributions are mapped onto the catalytic activity. The experimental dataset from Figure \ref{fig:Figure3}a is shown in this contour plot by black dots. The contour plot unveils the highest potential for mass activity improvement at nanoparticle diameters of 1 nm, 2 nm and 2.9 nm for nanoparticle size distributions below 0.2 nm. Those nanoparticles harbor mass activity enhancement of 152\%, 178\% and 190\% at (1.0$\pm$0.1) nm, (2.0$\pm$0.1) nm and (2.9$\pm$0.1) nm, respectively, compared to the highest experimental mass activity in Figure \ref{fig:Figure3}a. Recently realized elaborate fabrication methods enable such precise size control of Pt nanoparticle catalysts even down to the subnanometer scale \cite{Yamamoto2009c} giving rise to large catalytic activities at ($0.9\pm0.1$) nm nanoparticle size. This result corresponds perfectly with the computationally predicted activity peak at ($1.0\pm0.1$) nm in Figure \ref{fig:Figure4}. \\
To conclude, we have presented a computational model which enables rapid activity calculation of 3D Pt unstrained nanoparticle catalysts. In line with experiments, DFT studies show a linear scaling relation between OH-adsorption energies and generalized coordination numbers for Pt. We capitalize here on this crucial result to provide a link between the generalized coordination numbers and experimentally measured ORR catalytic activities. Making use of fundamental geometrical considerations, the presented computational model comprises the capability to determine nanoparticle mass activities in absolute units of $A/mg$, without the need for a reference to e.g. Pt(111). In this way, expensive DFT calculations are omitted during runtime realizing sharply reduced and feasible computation times in comparison to theoretical approaches which are based exclusively on DFT. The applicability of our computational model was tested on two experimental datasets involving particle size effects. Remarkably, the computational model accurately reproduces the experimental trends. Regarding the absolute units, the mass activities in both experiments differ considerably by one order of magnitude. Nonetheless, quantitative agreement in absolute units has been precisely observed between the computational model and one experimental dataset. Besides the capability to capture experimental activities on a highly accurate level, this study gives rise to predictions beyond currently available experiments. Promising nanoparticles, which harbor high mass activities, are predicted for nanoparticles sizes near 1 nm, 2 nm and 3 nm with size distributions below 0.2 nm. It is important to note that this complete nanoparticle size effect study was carried out within only few hours by means of the presented computational model. Thus, we believe that rapid nanoparticle activity calculation paves the way for high-throughput nanoparticle activity screening, which may strongly expedite the search for innovative catalysts in future studies.

\begin{acknowledgement}

This work is supported by the German
Research Foundation (DFG) under Grant No. 355784621 and the excellence cluster Nanosystems Initiative Munich (NIM) of the DFG. FCV thanks Spanish MEC for a Ram\'on y Cajal research contract (RYC-2015-18996).

\end{acknowledgement}

\begin{suppinfo}

{\bf Supporting Information.} The Supporting Information includes details of the DFT calculations, shapes of spherical nanoparticle catalysts, details of the computational model, and a showcase for the prediction of catalytic mass activities. 

\section{Details of the DFT calculations}

The DFT total energies to make Figure \ref{fig:Figure1}a are the following \cite{Calle-Vallejo2015}: the simulations were carried out using VASP \cite{Kresse1996}, the projector augmented-wave (PAW) method \cite{Joubert1999} and the PBE exchange-correlation functional \cite{Perdew1996}. For the relaxation of nanoparticles, all atoms were fully relaxed. For extended surfaces we used slabs with four metal layers: the topmost two and the adsorbates were allowed fully relaxed, while the bottommost two were fixed at the optimal bulk positions, found for PBE when the Pt-Pt distance is $2.81~\text{\AA}$. Cavities were simulated with 5-layer slabs, with the topmost three relaxed and the bottommost two fixed. The calculations were made with a plane-wave cut-off of $400~eV$ for nanoparticles and $450~eV$ for extended surfaces, using the conjugate-gradient scheme until the maximum force on any atom was below $0.01~eV~\text{\AA}^{-1}$. We used only the gamma point distribution for nanoparticles, whereas the k-point samplings for extended surfaces appear elsewhere \cite{Calle-Vallejo2014g}. The vacuum layer between periodically repeated images in extended surfaces was larger than $14~\text{\AA}$ and dipole corrections were included. Nanoparticles were calculated without dipole corrections in cubic boxes in which the shortest average distance between periodically repeated images was $\sim10~\text{\AA}$. We used $k_B T = 0.2 eV$ for the slab and nanoparticle calculations, and extrapolated the energies at $T=0K$. The DFT adsorption energies of $^*$OH were calculated as 
\begin{equation}
\Delta E_{OH} = E_{^*OH}-E_*-E_{OH} ~,
\end{equation}
where $^*$ is a free adsorption site. The gas-phase reference (OH) was calculated in cubic boxes of $3375~\text{\AA}^3$ using the gamma point only and $k_B T = 0.001~eV$.

\section{Spherical Nanoparticle Catalysts}

In this study, nanoparticle catalysts of quasi-spherical shape are investigated. Besides the nanoparticles with relatively small diameters presented in Figure \ref{fig:Figure1}b, we exemplify additional nanoparticles in Figure \ref{fig:FigureS1} featuring diameters in broadened range from 4 nm to 11 nm. 

\begin{figure}[H]
\centering
\includegraphics[width=\textwidth]{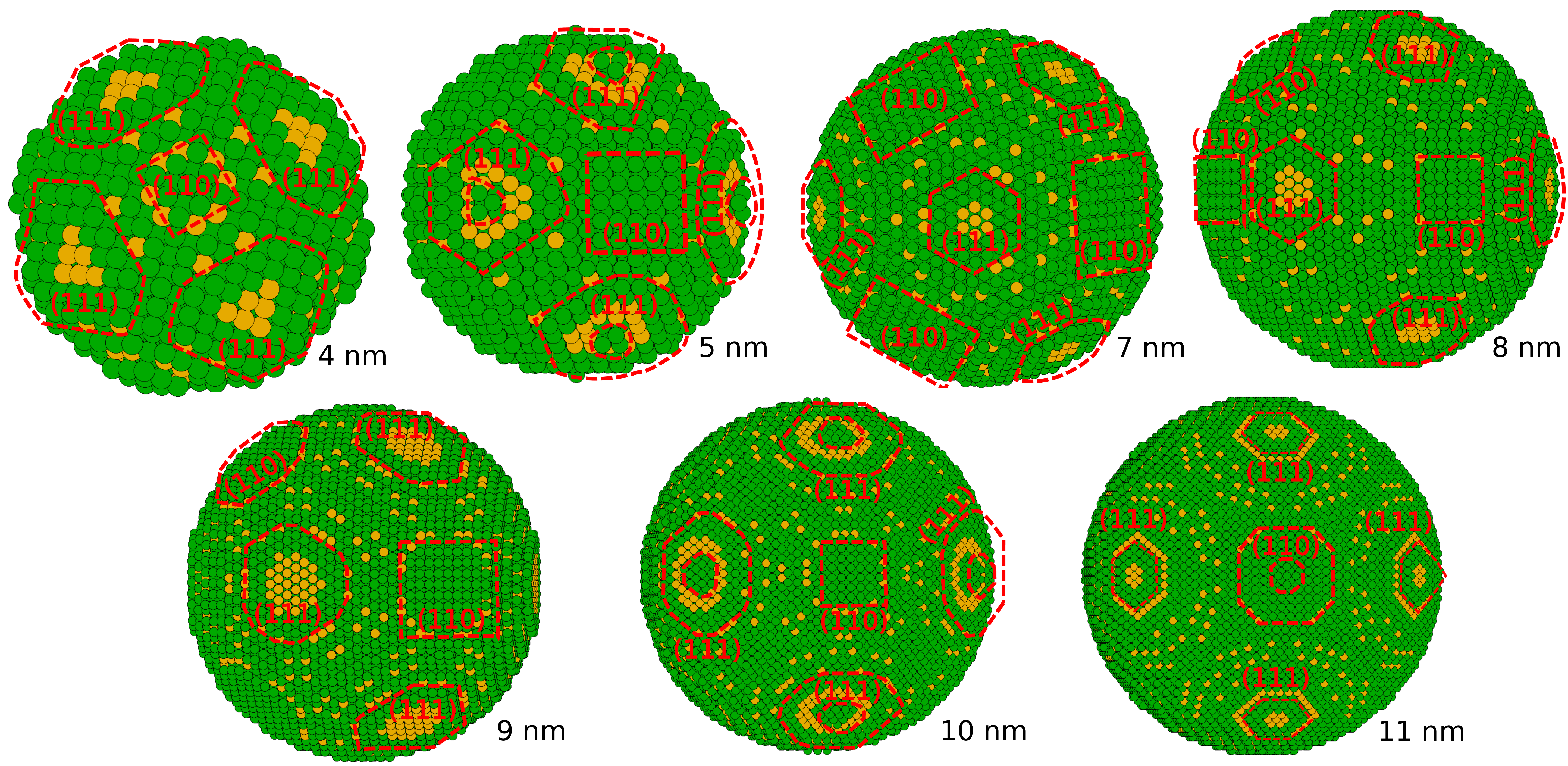}
\caption{Spherical nanoparticles are exemplified (in addition to those in Figure \ref{fig:Figure1}b) as they are examined in this study. The associated nanoparticle diameters range from 4 nm to 11 nm. Active sites with $7.5\leq\overline{CN}\leq8.3$ are highlighted in yellow. Prominent low index surfaces are enclosed by dashed lines. Nanoparticles are printed by ASE \cite{Larsen2017}.}
\label{fig:FigureS1}
\end{figure}

\section{Computational model}

In this section, we present the setup of our computational model in more details. First, we focus on the derivation of the fundamental volcano between experimental activities and $\overline{CN}$ in Figure \ref{fig:Figure1}d. Subsequently, we address the methodology to compute catalytic activities and mass activities in absolute units of $A$ and $A/mg_{Pt}$, respectively. We use the Atomic Simulation Environment (ASE) \cite{Larsen2017} for nanoparticle visualization and for calculations within the computational model.\\
The DFT analysis presented in Figure \ref{fig:Figure1}a yields the linear relation
\begin{equation}
\Delta E_{OH} = 0.1916 ~\overline{CN}-3.8673
\label{eqn:LinearRelationFigure1a}
\end{equation}
between OH-adsorption energies relative to OH in the gas phase and the generalized coordinatiuon number (gCN) $\overline{CN}$. In addition, experimental activities versus experimental OH-binding energies (given relatively to the OH-binding energy on the Pt(111) surface) are provided from literature and presented in Figure \ref{fig:Figure1}c. Thus, the linear DFT relation in Eq. \ref{eqn:LinearRelationFigure1a} is used to map the relative experimental OH-binding energies in Figure \ref{fig:Figure1}c onto $\overline{CN}$ as
\begin{equation}
\overline{CN} = \frac{\Delta E_{OH}-\Delta E^{Pt(111)}_{OH}}{0.1916}+\overline{CN}_{Pt(111)} ~.
\end{equation}
The scale of $\overline{CN}$ is appropriately renormalized by the gCN of surface atoms on Pt(111) which is given as $\overline{CN}_{Pt(111)}=7.5$. The resultant volcano plot, showing experimental activities versus $\overline{CN}$, is presented in Figure \ref{fig:Figure1}d. \\
The associated volcano-shaped activity trend is fitted by the function 
\begin{equation}
A_{volc}(\overline{CN}) =
\begin{cases}
A_1(\overline{CN}), \text{~~if~} \overline{CN}\leq \overline{CN}_p \\
A_2(\overline{CN}), \text{~~if~} \overline{CN}> \overline{CN}_p
\end{cases}
\label{eqn:A_volc}
\end{equation}
which consists of an increasing and a decreasing exponential function
\begin{align}
A_1(\overline{CN}) &= \text{exp}(a~\overline{CN}+b) ~, \\
A_2(\overline{CN}) &= \text{exp}(c~\overline{CN}+d) ~,
\end{align}
where $a>0$ and $c<0$. The peak of the volcano is given by the intersection of the functions $A_1$ and $A_2$ at $\overline{CN}_p$. Therefore, five fit parameters $a$, $b$, $c$, $d$ and $\overline{CN}_p$ are taken into account. The fit yields the functions 
\begin{align}
A_1(\overline{CN}) &= \text{exp}(3.14~\overline{CN}-23.40) ~, \\
A_2(\overline{CN}) &= \text{exp}(-4.96~\overline{CN}+42.18) ~,
\end{align}
which are provided in the inset in Figure \ref{fig:Figure1}d. The activity peak is given by the intersection of the functions at $\overline{CN}_p=8.1$ which corresponds to an OH-binding potential relative to Pt(111) of $\Delta E_{OH}-\Delta E^{Pt(111)}\approx 0.115~V$. \\
For activity calculation, the gCNs of all $N_{NP}$ nanoparticle atoms $i$ are evaluated. Using Eq. \ref{eqn:A_volc}, nanoparticle activities $j_{relNP}$, given relatively to the activity of one surface atom on Pt(111), are obtained by
\begin{equation}
j_{relNP} = \sum^{N_{NP}}_i \Theta\big(\overline{CN}(i)-7.5\big) \Theta\big(8.3-\overline{CN}(i)\big)~A_{volc}\big(\overline{CN}(i)\big)
\label{eqn:jrelNP}
\end{equation}
where $\overline{CN}(i)$ denotes the gCN at nanoparticle atom $i$ according to the definition in Eq. \ref{eqn:gCN}. The two involved Heaviside step functions
\begin{equation}
\Theta(x) = 
\begin{cases}
0, \text{~~if~} x<0 \\
1, \text{~~if~} x\geq0
\end{cases}
\end{equation}
set the activity contribution of nanoparticle sites with $\overline{CN}(i)<7.5$ or $\overline{CN}(i)>8.3$ to zero as discussed in the text. Such far, catalytic activities $j_{relNP}$ relative to Pt(111) are introduced. In the remainder of this section, we discuss the methodology to yield activities and mass activities in absolute units instead. \\
This approach is based on the experimental insight where the specific activity of the Pt(111) surface, expressed by the kinetic current density, has been measured to yield $j_{Pt(111)} = 2~mA/cm^2_{Pt}$ \cite{Calle-Vallejo2015}. The density of Pt atoms on the Pt(111) surface is given by \mbox{$d_{Pt(111)} = 1.503\text{~x~}10^{15}~cm^{-2}$}. Thus, employing the above-discussed relative catalytic activities $j_{relNP}$, absolute catalytic activities (expressed in absolute units of $A$) are calculated by 
\begin{equation}
j_{NP} = \frac{j_{Pt(111)}}{d_{Pt(111)}} j_{relNP} \approx 1.331\times10^{-18}~A ~j_{relNP} ~.
\end{equation}
The mass of one Pt atom is given by \mbox{$m_{Pt} = 195.084~u = 195.084 \text{~x~}1.661 * 10^{-21}~mg$}. Hence, catalytic mass activities are obtained by 
\begin{equation}
j_{mNP} = \frac{j_{NP}}{m_{Pt}N_{NP}} \approx \frac{4.107~A/mg_{Pt}}{N_{NP}} j_{relNP}
\label{eqn:jmNP}
\end{equation}
in absolute units of $A/mg_{Pt}$. 

\subsection{Showcase}

Here, we outline the above introduced methodology of our computational model by explicitly calculating the mass activity of the nanoparticle with 5 nm diameter. This nanoparticle is illustrated in Figure \ref{fig:FigureS2} which comprises $N_{NP}=4321$ atoms in total. The generalized coordination numbers 7.5, 7.75 and 8 occur at 48 sites each and additional 24 sites have generalized coordination $\overline{CN}=8.25$. Those sites are highlighted by colors in Figure \ref{fig:FigureS2}. Thus, Eq. \ref{eqn:jrelNP}, which yields catalytic activities relative to one surface atom on Pt(111), can be calculated as 
\begin{align}
j_{relNP} &= 48 \times \big(A_{volc}(7.5) + A_{volc}(7.75) + A_{volc}(8)\big) + 24 \times A_{volc}(8.25) \\
&\approx 48 \times \big( 1.16 + 2.55 + 5.58 \big) + 24 \times 3.53 \\
&= 530.64 ~.
\end{align}
Catalytic mass activities in absolute units are obtained by Eq. \ref{eqn:jmNP}. Eventually, this yields the mass activity
\begin{align}
j_{mNP}&\approx\frac{4.107~A/mg_{Pt}}{4321}~530.64 \\
&\approx0.50~A/mg_{Pt}
\end{align}
for the nanoparticle with 5 nm diameter. 

\begin{figure}[H]
\centering
\includegraphics[width=0.4\textwidth]{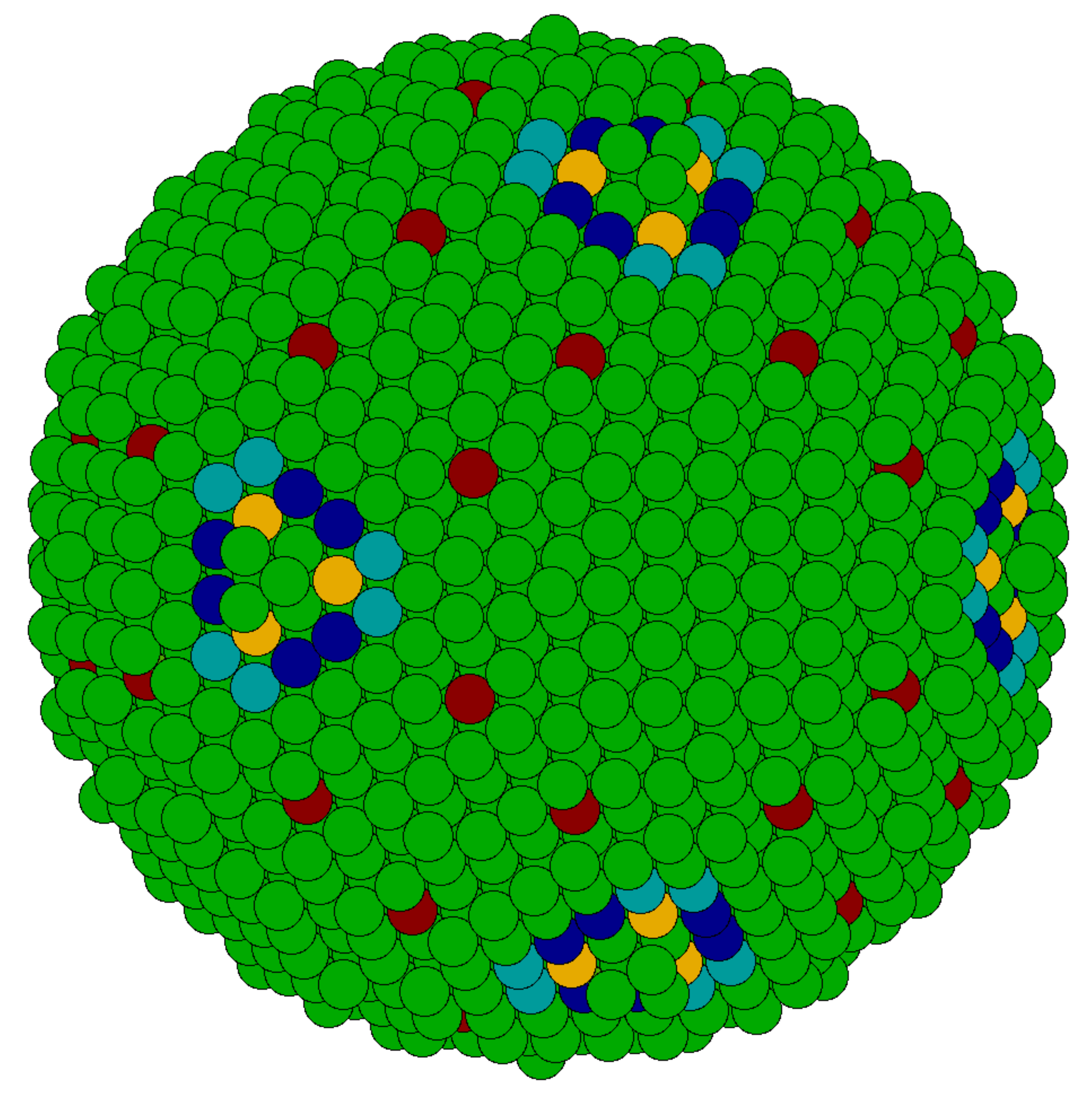}
\caption{Spherical nanoparticle with 5 nm diameter. Sites with $7.5\leq\overline{CN}\leq8.3$ are highlighted in light blue ($\overline{CN}=7.5$), dark blue ($\overline{CN}=7.75$), red ($\overline{CN}=8$) and yellow ($\overline{CN}=8.25$). The nanoparticle is printed by ASE \cite{Larsen2017}.}
\label{fig:FigureS2}
\end{figure}

\end{suppinfo}

\bibliography{achemso-demo}

\end{document}